\documentclass[sn-mathphys,Numbered]{sn-jnl}

\usepackage{graphicx}%
\usepackage{multirow}%
\usepackage{amsmath,amssymb,amsfonts}%
\usepackage{amsthm}%
\usepackage{mathrsfs}%
\usepackage[title]{appendix}%
\usepackage{xcolor}%
\usepackage{textcomp}%
\usepackage{manyfoot}%
\usepackage{booktabs}%
\usepackage{algorithm}%
\usepackage{algorithmicx}%
\usepackage{algpseudocode}%
\usepackage{listings}%

\usepackage[numbers]{natbib}
\begin{document}


\title[Article Title]{Results and Limits of Time Division Multiplexing for the BICEP Array High Frequency Receivers}


\author*[1]{\fnm{S.} \sur{Fatigoni}}\email{sofiaf@caltech.edu}

\author[2]{\fnm{P.A.R.} \sur{Ade}}
\author[3,4]{\fnm{Z.} \sur{Ahmed}}
\author[5]{\fnm{M.} \sur{Amiri}}
\author[6]{\fnm{D.} \sur{Barkats}}
\author[8,1]{\fnm{R.} \sur{Basu Thakur}}
\author[7]{\fnm{C.A.} \sur{Bischoff}}
\author[4]{\fnm{D.} \sur{Beck}}
\author[1,8]{\fnm{J.J.} \sur{Bock}}
\author[6]{\fnm{V.} \sur{Buza}}
\author[9]{\fnm{J.} \sur{Cheshire}}
\author[6]{\fnm{J.} \sur{Connors}}
\author[6]{\fnm{J.} \sur{Cornelison}}
\author[9]{\fnm{M.} \sur{Crumrine}}
\author[1]{\fnm{A.J.} \sur{Cukierman}}
\author[13]{\fnm{E.V.} \sur{Denison}}
\author[6]{\fnm{M.I.} \sur{Dierickx}}
\author[10]{\fnm{L.} \sur{Duband}}
\author[6]{\fnm{M.} \sur{Eiben}}
\author[11,12]{\fnm{J.P.} \sur{Filippini}}
\author[4]{\fnm{A.} \sur{Fortes}}
\author[1]{\fnm{M.} \sur{Gao}}
\author[7]{\fnm{C.} \sur{Giannakopoulos}}
\author[4]{\fnm{N.} \sur{Goeckner-Wald}}
\author[3,4]{\fnm{D.C.} \sur{Goldfinger}}
\author[4]{\fnm{J.A.} \sur{Grayson}}
\author[6]{\fnm{P.K.} \sur{Grimes}}
\author[9]{\fnm{G.} \sur{Hall}}
\author[4]{\fnm{G.} \sur{Halal}}
\author[5]{\fnm{M.} \sur{Halpern}}
\author[7]{\fnm{E.} \sur{Hand}}
\author[6]{\fnm{S.A.} \sur{Harrison}}
\author[3,4]{\fnm{S.} \sur{Handerson}}
\author[8,1]{\fnm{S.R.} \sur{Hildebrandt}}
\author[13]{\fnm{G.C.} \sur{Hilton}}
\author[13]{\fnm{J.} \sur{Hubmayr}}
\author[1]{\fnm{H.} \sur{Hui}}
\author[3,4]{\fnm{K.D.} \sur{Irwin}}
\author[4]{\fnm{J.H.} \sur{Kang}}
\author[4]{\fnm{K.S.} \sur{Karkare}}
\author[1]{\fnm{S.} \sur{Kefeli}}
\author[6]{\fnm{J.M.} \sur{Kovac}}
\author[3,4]{\fnm{C.L.} \sur{Kuo}}
\author[1]{\fnm{K.} \sur{Lau}}
\author[11]{\fnm{A.} \sur{Lennox}}
\author[4]{\fnm{T.} \sur{Liu}}
\author[8]{\fnm{K.G.} \sur{Megerian}}
\author[1]{\fnm{O.Y.} \sur{Miller}}
\author[1]{\fnm{L.} \sur{Minutolo}}
\author[1]{\fnm{L.} \sur{Moncelsi}}
\author[4]{\fnm{Y.} \sur{Nakato}}
\author[8]{\fnm{H.T.} \sur{Nguyen}}
\author[8,1]{\fnm{R.} \sur{O'Brient}}
\author[7]{\fnm{S.} \sur{Palladino}}
\author[6]{\fnm{M.A.} \sur{Petroff}}
\author[6]{\fnm{A.} \sur{Polish}}
\author[10]{\fnm{T.} \sur{Prouve}}
\author[9]{\fnm{C.} \sur{Pryke}}
\author[6]{\fnm{B.} \sur{Racine}}
\author[13]{\fnm{C.D.} \sur{Reintsema}}
\author[1]{\fnm{T.} \sur{Romand}}
\author[4]{\fnm{M.} \sur{Salatino}}
\author[1]{\fnm{A.} \sur{Schillaci}}
\author[6]{\fnm{B.L.} \sur{Schmitt}}
\author[9]{\fnm{B.} \sur{Singari}}
\author[8]{\fnm{A.} \sur{Soliman}}
\author[6]{\fnm{T.} \sur{St.Germaine}}
\author[1]{\fnm{A.} \sur{Steiger}}
\author[1]{\fnm{B.} \sur{Steinbach}}
\author[2]{\fnm{R.} \sur{Sudiwala}}
\author[3,4]{\fnm{K.L.} \sur{Thompson}}
\author[6]{\fnm{C.} \sur{Tsai}}
\author[2]{\fnm{C.} \sur{Tucker}}
\author[8]{\fnm{A.D.} \sur{Turner}}
\author[7]{\fnm{C.} \sur{Umiltà}}
\author[6]{\fnm{C.} \sur{Vèrges}}
\author[1]{\fnm{A.} \sur{Wandui}}
\author[8]{\fnm{A.C.} \sur{Weber}}
\author[5]{\fnm{D.V.} \sur{Wiebe}}
\author[9]{\fnm{J.} \sur{Willmert}}
\author[14]{\fnm{W.L.K.} \sur{Wu}}
\author[4]{\fnm{E.} \sur{Yang}}
\author[4]{\fnm{E.} \sur{Young}}
\author[4]{\fnm{C.} \sur{Yu}}
\author[6]{\fnm{L.} \sur{Zeng}}
\author[1]{\fnm{C.} \sur{Zhang}}
\author[1]{\fnm{S.} \sur{Zhang}}

\affil*[1]{\orgdiv{Department of Physics}, \orgname{California Institute of Technology}, \orgaddress{\city{Pasadena}, \postcode{91125}, \state{California}, \country{USA}}}

\affil[2]{\orgdiv{School of Physics and Astronomy}, \orgname{Cardiff University}, \orgaddress{\city{Cardiff}, \postcode{CF24 3AA}, \country{United Kingdom}}}

\affil[3]{\orgdiv{Kavli Institute for Particle Astrophysics and Cosmology}, \orgname{SLAC National Accelerator Laboratory}, \orgaddress{\street{2575 Sand Hill Rd}, \city{Menlo Park}, \postcode{94025}, \state{California}, \country{USA}}}

\affil[4]{\orgdiv{Department of Physics}, \orgname{Stanford University}, \orgaddress{ \city{Stanford}, \postcode{94305}, \state{California}, \country{USA}}}

\affil[5]{\orgdiv{Department of Physics and Astronomy}, \orgname{University of British Columbia}, \orgaddress{ \city{Vancouver}, \postcode{V6T1Z1}, \state{British Columbia}, \country{Canada}}}

\affil[6]{\orgdiv{Center for Astrophysics}, \orgname{Harvard and Smithsonian}, \orgaddress{\city{Cambridge}, \postcode{02138}, \state{Massachusetts}, \country{USA}}}

\affil[7]{\orgdiv{Department of Physics}, \orgname{University of Cincinnati}, \orgaddress{\city{Cincinnati}, \postcode{45221}, \state{Ohio}, \country{USA}}}

\affil[8]{\orgname{Jet Propulsion Laboratory}, \orgaddress{\city{Pasadena}, \postcode{91109}, \state{California}, \country{USA}}}

\affil[9]{\orgdiv{Minnesota Institute for Astrophysics}, \orgname{University of Minnesota}, \orgaddress{\city{Minneapolis}, \postcode{55455}, \state{Minnesota}, \country{USA}}}

\affil[10]{\orgdiv{Service des Basses Temperatures}, \orgname{Commissariat al Energie Atomique}, \orgaddress{\city{Grenoble}, \postcode{38054}, \country{France}}}

\affil[11]{\orgdiv{Department of Physics}, \orgname{University of Illinois at Urbana-Champaign}, \orgaddress{\city{Urbana}, \postcode{61801}, \state{Illinois}, \country{USA}}}

\affil[12]{\orgdiv{Department of Astronomy}, \orgname{University of Illinois at Urbana-Champaign}, \orgaddress{\city{Urbana}, \postcode{61801}, \state{Illinois}, \country{USA}}}

\affil[13]{\orgdiv{Kavli Institute for Cosmological Physics}, \orgname{University of Chicago}, \orgaddress{\city{Chicago}, \postcode{60637}, \state{Illinois}, \country{USA}}}

\affil[14]{\orgdiv{Department of Physics, Enrico Fermi Institute}, \orgname{University of Chicago}, \orgaddress{\city{Chicago}, \postcode{60637}, \state{Illinois}, \country{USA}}}

\affil[15]{\orgdiv{National Institute of Standards and Technology}, \orgaddress{\city{Boulder}, \postcode{80305}, \state{Colorado}, \country{USA}}}



\abstract{Time-Division Multiplexing is the readout architecture of choice for many ground and space experiments, as it is a very mature technology with proven outstanding low-frequency noise stability, which represents a central challenge in multiplexing. Once fully populated, each of the two BICEP Array high frequency receivers, observing at 150GHz and 220/270GHz, will have 7776 TES detectors tiled on the focal plane. The constraints set by these two receivers required a redesign of the warm readout electronics. The new version of the standard Multi Channel Electronics, developed and built at the University of British Columbia, is presented here for the first time.
BICEP Array operates Time Division Multiplexing readout technology to the limits of its capabilities in terms of multiplexing rate, noise and crosstalk, and applies them in rigorously demanding scientific application requiring extreme noise performance and systematic error control. 
Future experiments like CMB-S4 plan to use TES bolometers with Time Division/SQUID-based readout for an even larger number of detectors.}




\maketitle

\section{Introduction}\label{sec1}

\begin{figure}[H]
    \centering
    {\includegraphics[width=.6\textwidth]{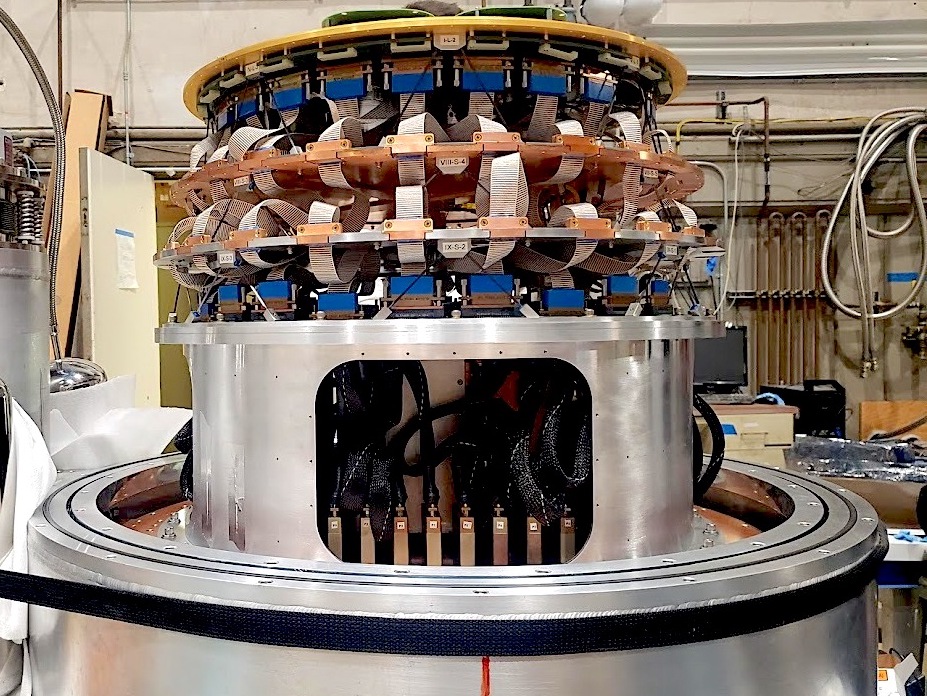}}
    \label{fig:ba2_readout}
    \caption{Picture of the sub-Kelvin stages of the BICEP Array 150GHz cryostat, showing the high density of readout cables that must go from 0.25 K to 4 K with TDM readout. The same number of cables runs from 4K to 300K in the lower part of the cryostat not visible in this picture.}
\end{figure}

\begin{figure}
    \centering
    {\includegraphics[width=.6\textwidth]{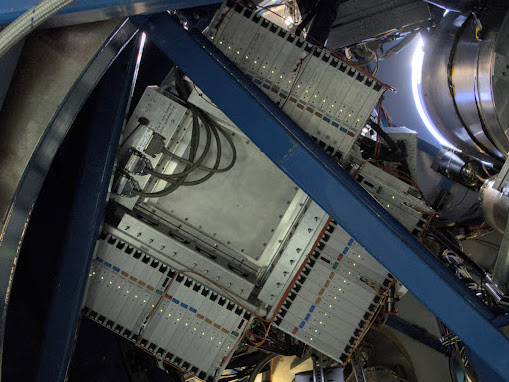}}
    \label{fig:dmce_on_cryo}
    \caption{The 3 Double MCEs mounted on the deployed BICEP Array 150GHz receiver at the South Pole.}
\end{figure}

In the Time Division Multiplexing (TDM) readout scheme, each superconducting bolometer is inductively coupled to a SQUID amplifier. The detectors are grouped by rows and columns, and the signal is read out from all columns simultaneously, turning on one row at a time.
Because the SQUID's response is intrinsically non-linear a feedback loop is used to calculate the correct flux to be sent to the SQUIDs feedback to compensate for changes in detector current and keep the amplification in a linear regime. This feedback current is also recorded as “signal”. The Multi Channel Electronics (MCE) system, developed and built at the University of British Columbia, supplies the bias current needed to keep the TES detectors in transition, controls the multiplexer and SQUID amplifiers and reads the signal from a $41 \times 32$ element array (\cite{battistelli2008mce}, \cite{Hasselfield_thesis_2013}).\newline
The total number of detectors that can be read out with TDM is limited by the thermal load induced by cables running from $250 mK$ to $300 K$ (Fig.\ref{fig:ba2_readout}), and the volume available to fit readout cables inside the cryostat and to fit warm electronics in the cryostat mount(Fig.\ref{fig:dmce_on_cryo}).
To address the latter problem, the BICEP Array (BA) two high frequency receivers use a new version of the MCE, the Double MCE (DMCE), that can read out a matrix of 41 rows x 64 columns (\cite{Fatigoni_2023}). 
Time Division Multiplexing is the baseline technology for CMB-S4 (\cite{barron2022conceptual}, \cite{Goldfinger23}).

\section{Detector Modules}\label{sec2}
\begin{figure}[H]
  \centering
\includegraphics[width=0.5\textwidth]{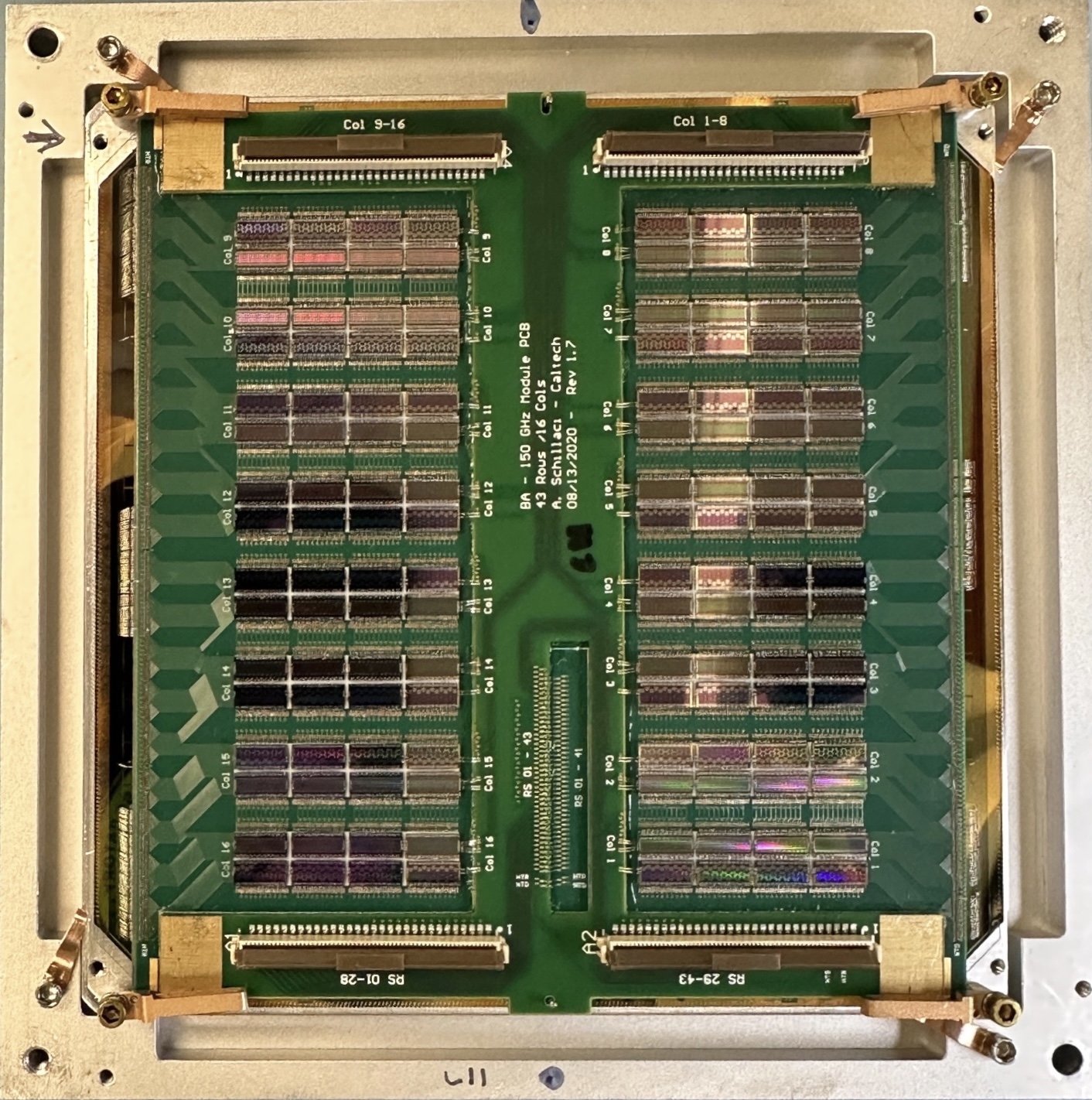}
 \caption[]{A detector module with exposed MUX Nyquist PCB, fully populated with SQUID chips. The picture shows the high density of lines that are routed from the chips to the sides of the PCB. The number of wirebonds is equally large. }
 \label{fig:module}
\end{figure}

The detector formats set up the requirements for the TDM readout electronics.
The detectors used in BICEP Array are photon-noise-limited Transition Edge Superconducting (TES) bolometers (\cite{Zhang_2020}).
The pixel count in BA detector tiles is limited by the receiver frequency (see Table \ref{tab:rxs}). The two BA high frequency receivers, observing at 150GHz and 220/270GHz, share the same detector count and detector distribution on a tile, which is a 18x18 pixel tile, for a total of 324 pixels, or 648 detectors. \newline
Detector tiles are included into a modular architecture which is shared in between all the BA receivers (\cite{schillaci2021bicep}).
Each module hosts SQUID chips for multiplexing, which interface with the detector tile through an interface PCB, to which they are individually wirebonded. The number of wirebonds required is one of the challenging factors in using TDM for high density pixels arrays (Fig.\ref{fig:module}). 
All the SQUID chips (MUX and Nyquist) are made at NIST. \newline
In 8 of the detectors in the matrix, the TES antenna is not connected, making these detectors sensitive to thermal drifts, RFI and direct island stimulation. Two other pixels are occupied by loss test devices, where the incoming power is split equally between two paths, one of which has a lossy meander of known length. These two devices are used to estimate the signal loss through the feedline.\newline
The detector module is composed of a layer stack that includes a quartz anti-reflection (AR) coating tile, the detector wafer, the $\lambda/4$ backshort, an A4K magnetic shield and two custom designed PCBs.
The module stack is housed in a superconducting niobium and aluminum box, that together with a high-$\mu$ A4K sheet inside the module, is designed to achieve high magnetic shielding performance (\cite{moncelsi2020receiver}).
The stack of layers inside each module is held together by custom designed berillium-copper clips that apply the right amount of pressure to hold the layers together.

\begin{center}
\begin{table}
\small
\centering
\begin{tabular}[c]{|l|l|l|l|l|}
\hline
Receiver       & Number of  & Mux       & MCE       & Number \\
Observing Band & Detectors per&   Factor        & Type      &       of\\
(GHz)          & Module/Receiver&   &    &   MCEs \\
\hline\hline   						
					
     &          &                      &            & \\
$\big \langle \hspace{-3pt}
\begin{array}{l} 30 \\ 40 \end{array}$ & $\begin{array}{r} 32/192 \\ 50/300 \end{array}$
  & $\begin{array}{r} 33 \\ 33 \end{array}$ & $\begin{array}{r}  Single \\  Single \end{array}$
  & $\begin{array}{r} 1 \\  \end{array}$       \\
 $\hphantom{\big \langle}  \hspace{-3pt}
\begin{array}{l} 95 \\ 150 \end{array}$   & $\begin{array}{r} 338/4,056 \\ 648/7,776 \end{array}$
  & $\begin{array}{r} 43 \\ 41 \end{array}$ & $\begin{array}{r}Single \\ Double \end{array}$
  & $\begin{array}{r} 3 \\ 3 \end{array}$       \\
$\big \langle \hspace{-3pt}
\begin{array}{c} 220 \\ 270 \end{array}$   & $\begin{array}{r} 648/7,776 \\ 648/7,776 \end{array}$
  & $\begin{array}{r} 41 \\ 41 \end{array}$ & $\begin{array}{r} Double \\ Double \end{array}$
  & $\begin{array}{r} 3 \\  \end{array}$       \\
\hline
\end{tabular}
\caption{Detector count, MUX Factor and Readout electronics needed for each of the four BICEP Array receivers.
Readout channels count for the three TDM BICEP Array receivers at 30/40, 95 and 150 GHz. BA1
30/40 GHz receiver is split in 2 colors counting each as a 0.5 MCE. }
\label{tab:rxs}
\end{table}

\end{center}


\section{Warm Readout Electronics}\label{sec3}
\begin{figure}[H]
  \centering
\includegraphics[width=0.7\textwidth]{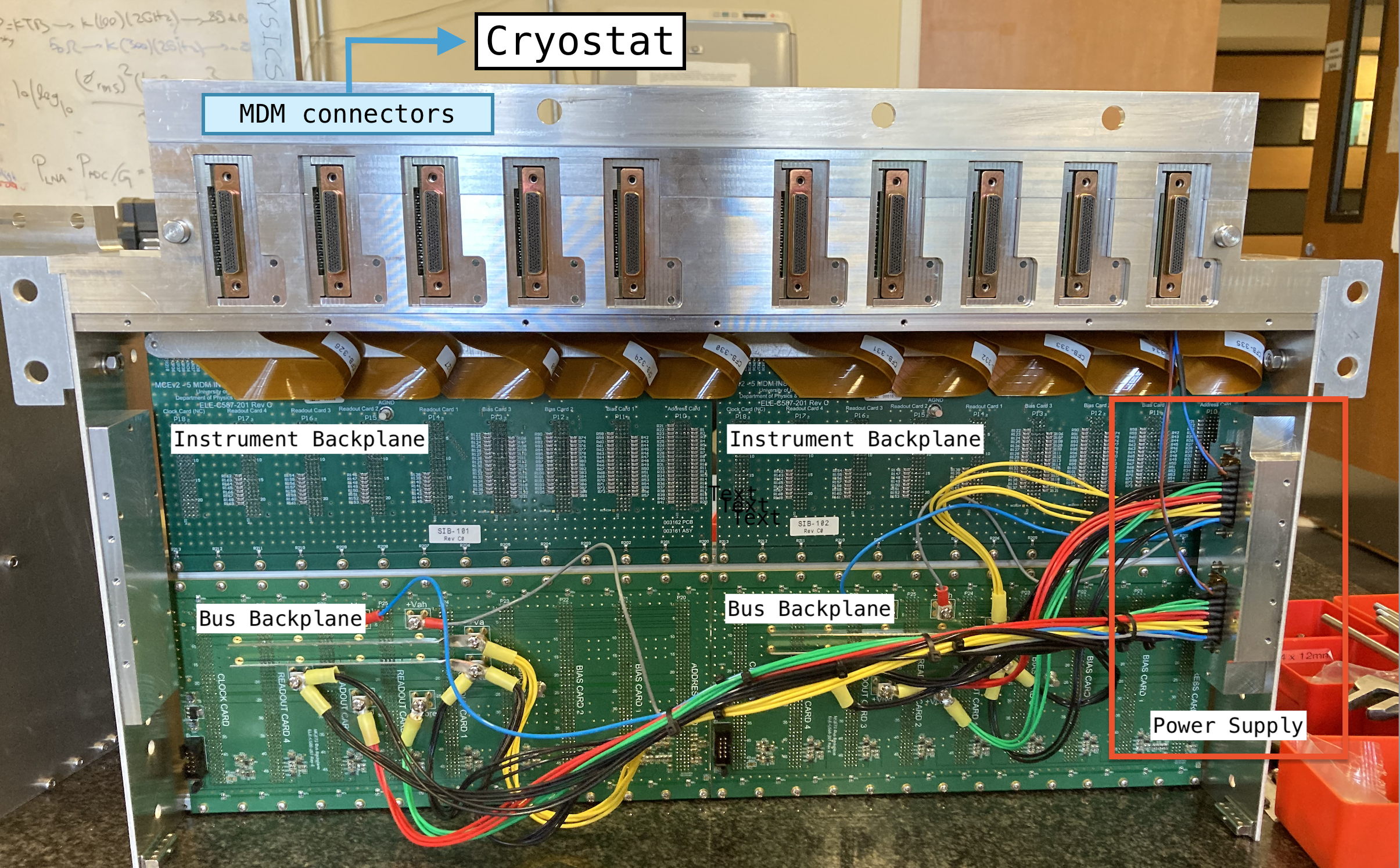}
 \caption[]{Picture of the DMCE side which mates to the cryostat. The back cover is removed in this picture. The filter boxes are inserted in a one piece aluminum block called filter rail, which is aligned to the cryostat through two alignment pins. MDM connectors can be translated in the direction orthogonal to this page turning a set of shafts, accessible from the opposite DMCE side with a torque ranch.}
 \label{fig:dmce2}
\end{figure}

The multiplexing is controlled at room temperature by the MCE (\cite{battistelli2008mce}), based on a modular architecture, where four different kind of cards are in charge of different operations.
The number of cards and MDM connectors in one MCE subrack varies and the required MCE version depends on the size of the array that needs to be read out.
For the BICEP Array 30/40GHz receiver, a standard MCE subrack is used. A standard MCE includes one clock card, three bias cards and either two or four readout cards.\newline 
For the BICEP Array 150GHz receiver, which incorporates $\sim 8000$ TESs in the focal plane, a mechanical redesign was required to double the number of detectors that a single MCE can read out.
The DMCE includes two clock cards, two address cards, eight readout cards and six bias cards.
The signal is routed through ten MDM connectors and to two bus backplanes and two instrument backplanes (see Fig.\ref{fig:dmce2}). The MDM connectors are encapsuled in aluminum enclosures called filter boxes.
In order to increase the linear density of connectors, in the new version of the MCEs, we rotated the filter boxes of 90 degrees with respect to the filter rail. Making the filter rail shorter than the rest of the subrack was necessary in order to fit the necessary 3 DMCE subracks on the bottom of the cryostat [Fig. \ref{fig:dmce_on_cryo}]. \newline
The flexes that carry the signal from the MDM connector to the backplanes are longer than in the previous version and need to be installed turning them in a specific twisted S shape before mating them to the Hirose connectors on the backplane (see Fig.\ref{fig:dmce2}).
The two bus backplanes are powered independently by two power connectors. The two power connectors are assembled together in a aluminum cassette and can be mounted on either side panel of the subrack. In the DMCE the fans are automatically powered by these same harness that powers the backplanes, and are turned on whenever the power it is connected.\newline
The mating between each DMCE subrack and the cryostat is done at the bottom of the cryostat, through a double density feedthrough flange. The two alignment pins at the extreme ends of the filter rail are mated to a pin and a slot on the feedthrough flange side. In order for the mating to be successful, the two pieces of each flange need to be aligned with high precision. To make sure that this is the case an alignment tool was built. The alignment tool is a single long piece of metal that grabs the two half flanges' pin holes and keeps the flanges aligned while they get mounted to the cryostat.\newline
Finally, the DMCE is attached to the cryostat through a long mounting flange located on top of the filter rail and two mounting ears per side. A DMCE can read out a matrix of 41 rows x 64 columns.\newline
DMCEs come from a mostly mechanical redesign of the standard MCE subrack, where we have not only made the crate smaller but also optimised the shape to fit a set of three crates on a single cryostat. The DMCE backplane schematics are similar to the previous version, with small changes in the signal assignment. A further reduction in size would not be possible without a major electrical redesign and signal rerouting.\newline
The DMCEs designed and built for the BA 150GHz receiver showed reliability and high performance. 

\section{Multiplexing Rate}\label{sec4}
\begin{figure}[H]
  \centering
 \includegraphics[width=0.7\textwidth]{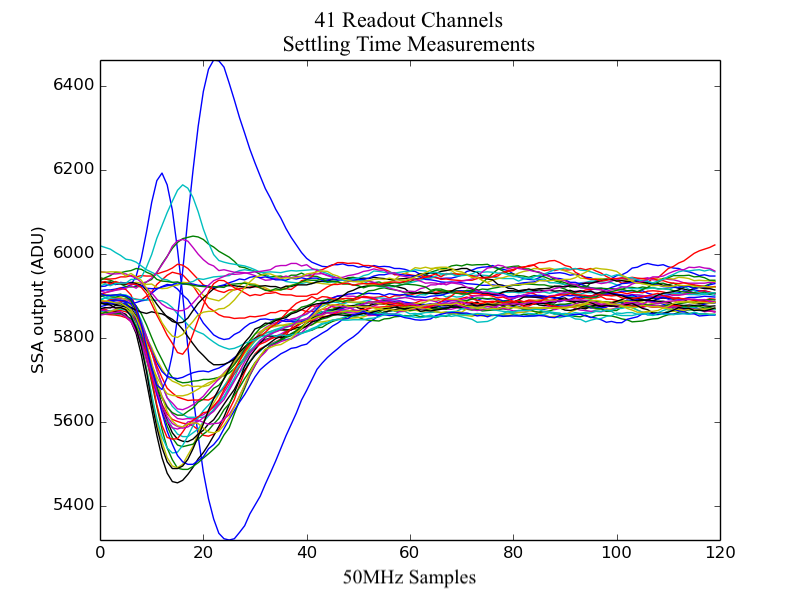}
  \caption[Timestreams showing the SQUIDs behaviour after switching to a new row]{Timestreams showing the SQUID behaviour after switching to a new row, on the BA 150 GHz receiver. These measurements were acquired as part of the BICEP Array 150GHz receiver readout characterization. This plot shows timestreams for all rows on one column overlapped.
  The data set was acquired at a sampling frequency of $f_{\text{samp}}=50MHz$, switching between rows every 120 clock cycles.}
  \label{fig:SQ_sett_time1}
\end{figure}

\begin{figure}[H]
  \centering
    \includegraphics[width=0.7\textwidth]{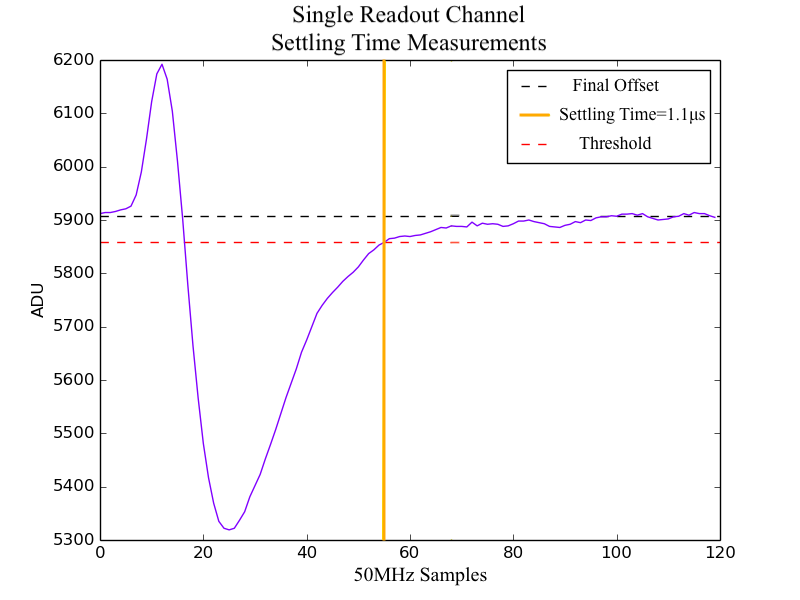}
  \caption[SQUID chain settling behaviour for a single readout channel]{This plot shows a single timestream for one row in the column (that is one SQUID+TES pair) picked from the ensemble in Fig.\ref{fig:SQ_sett_time1}.
  Here is plotted the final offset level, the threshold used to extract the settling time ($offset\pm50ADU$) and the resulting settling time for this single readout channel.
}
  \label{fig:SQ_sett_time2}
\end{figure}

The time that we spend on each row when multiplexing, which we call row length, needs to be tuned.
Ideally we want to multiplex as fast as we can (i.e. making the length of each row visit to be as short as possible), so as to limit the contribution to total noise from SQUID aliased noise. However, in practice the multiplexing rate is limited by the SQUID settling time.
Measurements of the SQUID settling time are acquired as part of the readout characterization process. The goal is to acquire a direct measurement of the minimum time interval the electronics need to spend on each. This parameter is extracted as the time that it takes for the SQ1 to go back within some threshold of its final offset value, after which we established the SQUID converges quickly to the final offset we are reading out. This data set is acquired sampling at a frequency of $f_{\text{samp}}=50MHz$.\newline
For the BICEP Array 150GHz receiver, measurements of the settling time show that we can set the length of the row visit as short as $\sim 1 \mu s$ (see Fig.\ref{fig:SQ_sett_time1} and Fig.\ref{fig:SQ_sett_time2}).

\begin{figure}[H]
  \centering
\includegraphics[width=0.8\textwidth]{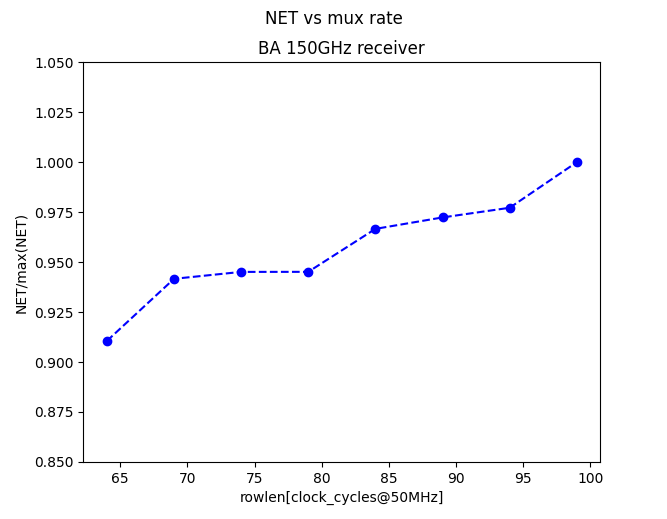}
 \caption[]{Noise Equivalent Temperature (NET) as a function of switching rate, normalized to the value with a $2 \mu s$ switching rate (100 clock cycles at 50MHz). as a function of the number of clock cycles we wait when switching rows. This data set was acquired with the fielded BICEP Array 150GHz receiver. This data set shows that pushing down the row length from 100 to 65 results in a $\sim 9\%$ improvement in NET. }
 \label{fig:NET}
\end{figure}

Additionally, for higher frequency receivers the number of rows we want to multiplex (MUX factor) is higher (see Tab.\ref{tab:rxs}).
In the BICEP Array 30/40GHz receiver we multiplex 33 rows, while in the 150GHz receiver we multiplex 41 rows. \newline
The multiplexing rate is given by MUX rate = switch rate / number of rows, where the switch rate is the frequency at which we switch between rows when multiplexing.
If the number of multiplexed rows increases at constant switch rate, our mux rate decreases and we pay a noise penalty from introducing extra aliased noise into our dataset. \newline
As a matter of fact, we measured the 90th percentile of the SQUID settling time to be $1.3 \mu s$(65 clock cycles) in the BA 150GHz receiver, and  $2 \mu s$ (100 clock cycles) in the BA 30/40GHz receiver.
By increasing the switching rate accordingly, we measured a $\sim 10\%$ improvement in Noise Equivalent Temperature (NET) in the BA 150GHz receiver (see Fig.\ref{fig:NET}).

\section{Conclusion}

Time Domain Multiplexing is a mature and well-characterized technology, with a noise performance that is suitable for high demanding applications aiming to set tight cosmological constraints(\cite{BK18}). In all BICEP/Keck telescopes, MCEs have been in charge of the Time Domain Multiplexing. 
The detector density in the BICEP Array high-frequency receivers required a redesign of the readout electronics to double the number of channels that a single subrack can read out. The Double MCEs have been deployed to the South Pole and used to characterize the readout performance of the BICEP Array 150GHz receiver. For the BA 150GHz receiver, we were able to increase our multiplexing rate by 35\%, which resulted in a 9\% improvement in the detector NETs.

\section*{Acknowledgement}
BICEP/Keck Array project have been made possible through a series of grants from the National Science Foundation and by the Keck Foundation. The development of antenna-coupled detector technology was supported by the JPL Research and Technology Development Fund and NASA Grants. The development and testing of focal planes were supported by the Gordon and Betty Moore Foundation at Caltech. Readout electronics were supported by a Canada Foundation for Innovation grant to UBC. The computations in this paper were run on the Odyssey cluster supported by the FAS Science Division Research Computing Group at Harvard University. We thank the staff of the U.S. Antarctic Program and in particular the South Pole Station without whose help this research would not have been possible. Tireless administrative support was provided by Nancy Roth-Rappard.

\backmatter

\bibliography{sn-bibliography.bib}

\end{document}